\documentclass[pre,aps,twocolumn,superscriptaddress,showpacs]{revtex4-1}
\usepackage{amssymb}
\usepackage{epsfig}
\usepackage{amsmath}
\usepackage{subfigure}
\usepackage{graphicx}
\usepackage{textcomp}
\usepackage{url}
\usepackage{float}
\usepackage{color}
\usepackage{cases}
\usepackage[normalem]{ulem}

\usepackage[colorlinks=true, urlcolor=blue, anchorcolor=blue, citecolor=blue,filecolor=blue,linkcolor=blue,menucolor=blue
]{hyperref}

\usepackage{color}

\usepackage{epsfig}
\usepackage{textcomp}
\usepackage{epstopdf}

\newcommand{\antiquad}{\!\!\!\!\!\!\!\!}

\begin{document}
\title{Exact short-time height distribution for the flat Kardar-Parisi-Zhang interface}

\author{Naftali R. Smith}
\email{naftali.smith@mail.huji.ac.il}
\author{Baruch Meerson}
\email{meerson@mail.huji.ac.il}
\affiliation{Racah Institute of Physics, Hebrew University of
Jerusalem, Jerusalem 91904, Israel}


\begin{abstract}

We determine the exact short-time distribution
$-\ln \mathcal{P}_{\text{f}}\left(H,t\right)= S_{\text{f}} \left(H\right)/\sqrt{t}$ of the one-point height $H=h(x=0,t)$ of an evolving 1+1 Kardar-Parisi-Zhang (KPZ) interface for flat initial condition.
This is achieved by combining (i) the optimal fluctuation method, (ii) a time-reversal symmetry of the KPZ equation in 1+1 dimension, and (iii) the recently determined exact short-time height distribution $-\ln \mathcal{P}_{\text{st}}\left(H,t\right)= S_{\text{st}} \left(H\right)/\sqrt{t}$ for \emph{stationary} initial condition. In studying the large-deviation function $S_{\text{st}} \left(H\right)$ of the latter, one encounters two branches: an analytic and a non-analytic. The analytic branch is non-physical beyond a critical value of $H$ where a second-order dynamical phase transition occurs. Here we show that, remarkably, it is the analytic branch of  $S_{\text{st}} \left(H\right)$ which determines the large-deviation function $S_{\text{f}} \left(H\right)$ of the flat interface via a simple mapping
$S_{\text{f}}\left(H\right)=2^{-3/2}S_{\text{st}}\left(2H\right)$.

\end{abstract}

\maketitle

\section{Introduction}

The one-dimensional Kardar-Parisi-Zhang (KPZ) equation \citep{KPZ} describes the non-equilibrium stochastic dynamics of the height $h(x,t)$ of a growing interface at the point $x$ of a substrate at time $t$ as
\begin{equation}
\label{eq:KPZ_dimensional}
\partial_{t}h=\nu\partial_{x}^{2}h+\frac{\lambda}{2}\left(\partial_{x}h\right)^{2}+\sqrt{D}\,\xi(x,t).
\end{equation}
Here $\xi(x,t)$ is a Gaussian noise with zero average
and
\begin{equation}\label{correlator}
\langle\xi(x_{1},t_{1})\xi(x_{2},t_{2})\rangle = \delta(x_{1}-x_{2})\delta(t_{1}-t_{2}).
\end{equation}
At late times, the lateral correlation length grows as $t^{2/3}$, and the interface width grows as $t^{1/3}$.
The famous exponents $2/3$ and $1/3$ define an important universality class of non-equilibrium growth \cite{HHZ,Barabasi,Krug,Corwin,QS,S2016,Takeuchi2017}.

In recent years, much effort has gone into the study of quantities which describe the height fluctuations of the KPZ interface in greater detail. One of them is the full probability distribution $\mathcal{P}\left(H,t\right)$ of the interface height at specified point and time $H=h\left(x=0,t\right)$. The shape of this distribution strongly depends  the initial condition $h\left(x,t=0\right)$, and this dependence persists at arbitrarily long times \cite{QS,S2016,Takeuchi2017}.
The three most physically relevant, and extensively studied initial conditions are the sharp-wedge or ``droplet'', where $h\left(x,t=0\right)=\left|x\right|/\delta$ with $\delta \to 0$, the flat initial condition $h\left(x,t=0\right)=0$ and the ``stationary'' interface, where the initial height profile $h(x,t=0)$ is randomly chosen from the steady-state distribution.
A significant achievement was the derivation of exact representations for a generating function of $\exp[(\lambda/2\nu)H]$, at any time, for these three initial conditions \citep{SS,CDR,Dotsenko,ACQ, CLD,IS,Borodinetal}. 
These representations are given in terms of Fredholm determinants or Pfaffians.
Using the exact representations, it was shown that typical fluctuations at long times, $t\gg \nu^{5}/(D^{2}\lambda^{4})$,
for the flat, ``droplet'' and stationary initial conditions are described by the Gaussian orthogonal ensemble (GOE) Tracy-Widom distribution \citep{TracyWidom1996}, the Gaussian unitary ensemble (GUE) Tracy-Widom distribution and the Baik-Rains distribution \citep{BR}, respectively.

Lately, there has been growing interest in \emph{short-time} $t \ll \nu^{5}/(D^{2}\lambda^{4})$ fluctuations of the KPZ interface height. Here the \emph{tails} of $\mathcal{P}\left(H,t\right)$ were seen to exhibit new scaling behavior. A direct method of obtaining the short-time distribution $\mathcal{P}\left(H,t\right)$ is by extracting the  short-time asymptotics from the exact representations. These calculations, although technically difficult, were performed for the droplet \citep{DMRS} and stationary \citep{LeDoussal2017} initial conditions. In both cases, the distributions scale, in a proper moving frame \citep{footnote:displacement}, as $-\ln\mathcal{P}\simeq S \left(H\right)/\sqrt{t}$, where the large-deviation functions $S(H)$ were found exactly.
For the flat initial condition, no such calculation has yet been performed, but the first four cumulants of the distribution were found using the Fredholm representations \citep{Gueudre}.

A powerful, although approximate, alternative tool for studying height fluctuations is the optimal fluctuation method (OFM), also known by the names weak-noise theory, instanton method and macroscopic fluctuation theory. Originating in condensed matter physics \cite{Halperin,Langer,Lifshitz,Lifshitz1988},   the OFM  found applications in the studies of turbulence and turbulent transport
\citep{turb1,turb2,turb3}, diffusive lattice gases \citep{bertini2015} and
stochastic reactions on lattices \citep{EK, MS2011}.
It was applied to the KPZ equation and closely related systems in Refs. \citep{Mikhailov1991, GurarieMigdal1996,Fogedby1998, Fogedby1999,Nakao2003, KK2007,KK2008,KK2009,Fogedby2009,MKV,KMSparabola,Janas2016,
MeersonSchmidt2017, MSV_3d, SMS2018, SKM2018}.
In the OFM the path integral of the stochastic process, conditioned on a specified large deviation, is evaluated using the saddle-point approximation. Of course, this requires a proper small parameter.
The ensuing minimization procedure leads to equations for the optimal (most likely) path of the system and the most likely realization of the noise, which can be cast into Hamiltonian form. The ``classical'' action, evaluated on the optimal path, yields $-\ln\mathcal{P}$ up to a pre-exponential factor.
Importantly for the following, for the KPZ equation in 1+1 dimension, the OFM becomes asymptotically exact in the short-time limit $t \to 0$.

The OFM immediately yields the short-time scaling behavior $-\ln\mathcal{P}\simeq S \left(H\right)/\sqrt{t}$.
The large-deviation function $S \left(H\right)$ is found from the solution of the ensuing variational problem. The latter involves nonlinear and coupled partial differential equations, so this solution is difficult to obtain.
To date, no exact analytic solution to the OFM's variational problem for the KPZ equation has been found for arbitrary $H$. Approximate analytical solutions for small $H$, for large positive $H$ and for large negative $H$ were obtained
for the flat \citep{KK2007,KK2008,KK2009,MKV}, ``droplet'' \citep {KMSparabola} and stationary \citep{Janas2016} initial conditions. For the flat interface, $S(H)$ was computed numerically in a broad range of $H$ \cite{MKV}.

As regards the large-deviation function $S_{\text{st}} \left(H\right)$ for the stationary interface, the OFM led to the discovery of a second-order dynamical phase transition: a jump of the second derivative $\partial_{H}^{2}S_{\text{st}}$ at a critical value of $\lambda H=\lambda H_c>0$.
This phase transition is due to a spontaneous breaking of the spatial mirror symmetry $x \leftrightarrow -x$ of the optimal path of the  interface \cite{Janas2016}. In addition to the non-analytic branch of $S_{\text{st}}(H)$, there is an analytic branch, corresponding to spatially-symmetric interface histories. At $\lambda H > \lambda H_c$, this branch gives a larger action, so it is not optimal and therefore not physical \citep{Janas2016}. Both of the branches of $S_{\text{st}}(H)$ were subsequently found exactly by
Krajenbrink and Le Doussal \cite{LeDoussal2017} from the exact representations. Recently, a Landau theory was developed
for this dynamical phase transition \cite{SKM2018}.

In this work we use the OFM in order to obtain the exact analytical form of the short-time large-deviation function $S_{\text{f}}(H)$ for the flat initial condition, see Fig.~\ref{fig:sF} and Eqs.~(\ref{eq:Psi_def})-(\ref{eq:sF_exact}) below.
We achieve this without solving the OFM problem explicitly.
Rather, we incorporate into the OFM formalism a recently established time-reversal symmetry of the KPZ equation. Then we exploit this symmetry to obtain an exact mapping between the OFM formulations for the flat and stationary KPZ interfaces.
This enables us to obtain the remarkably simple relation
\begin{equation}\label{mainresult}
S_{\text{f}} (H)= \frac{1}{2 \sqrt{2}} S_{\text{st}} (2H)
\end{equation}
between $S_{\text{f}}(H)$ and the \emph{analytic} branch of $S_{\text{st}}(H)$. As this branch is  known exactly \cite{LeDoussal2017}, we immediately obtain exact $S_{\text{f}}(H)$, which is the main result of this work.

\begin{figure}[ht]
\includegraphics[width=0.45\textwidth,clip=]{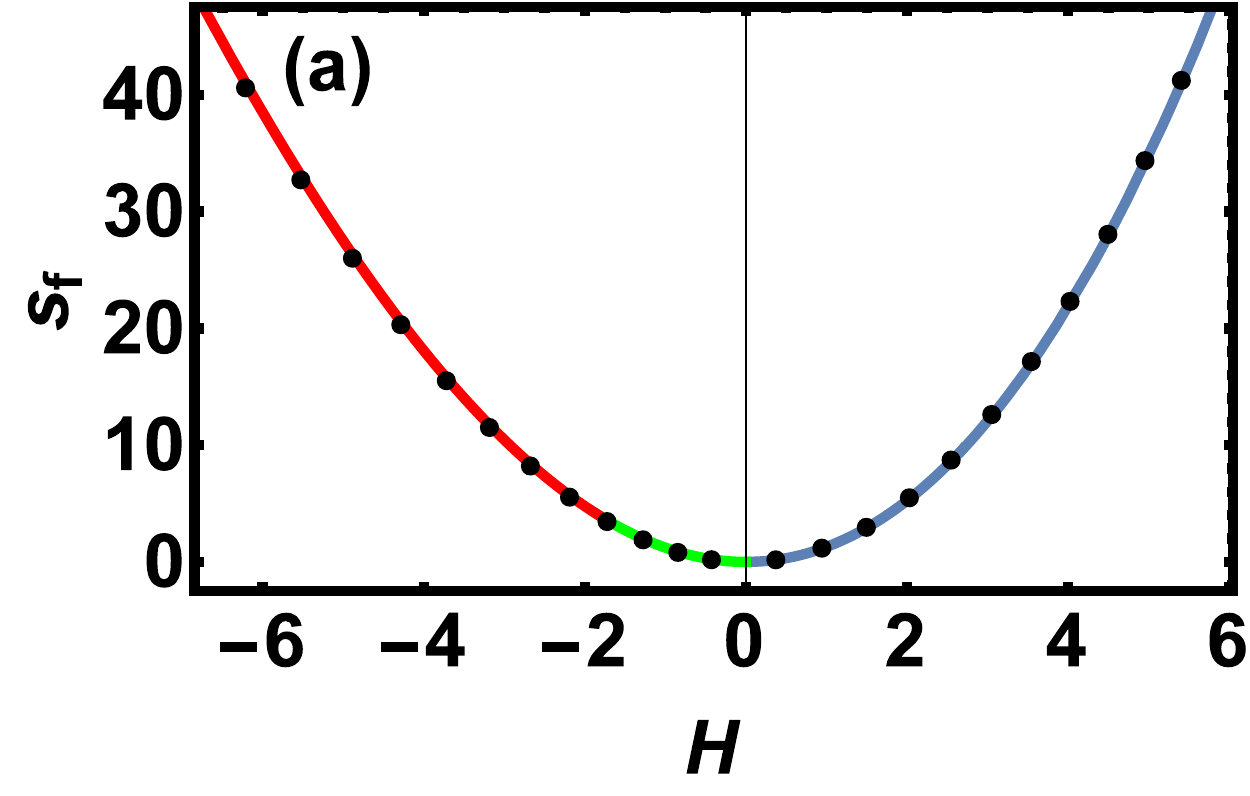}
\includegraphics[width=0.45\textwidth,clip=]{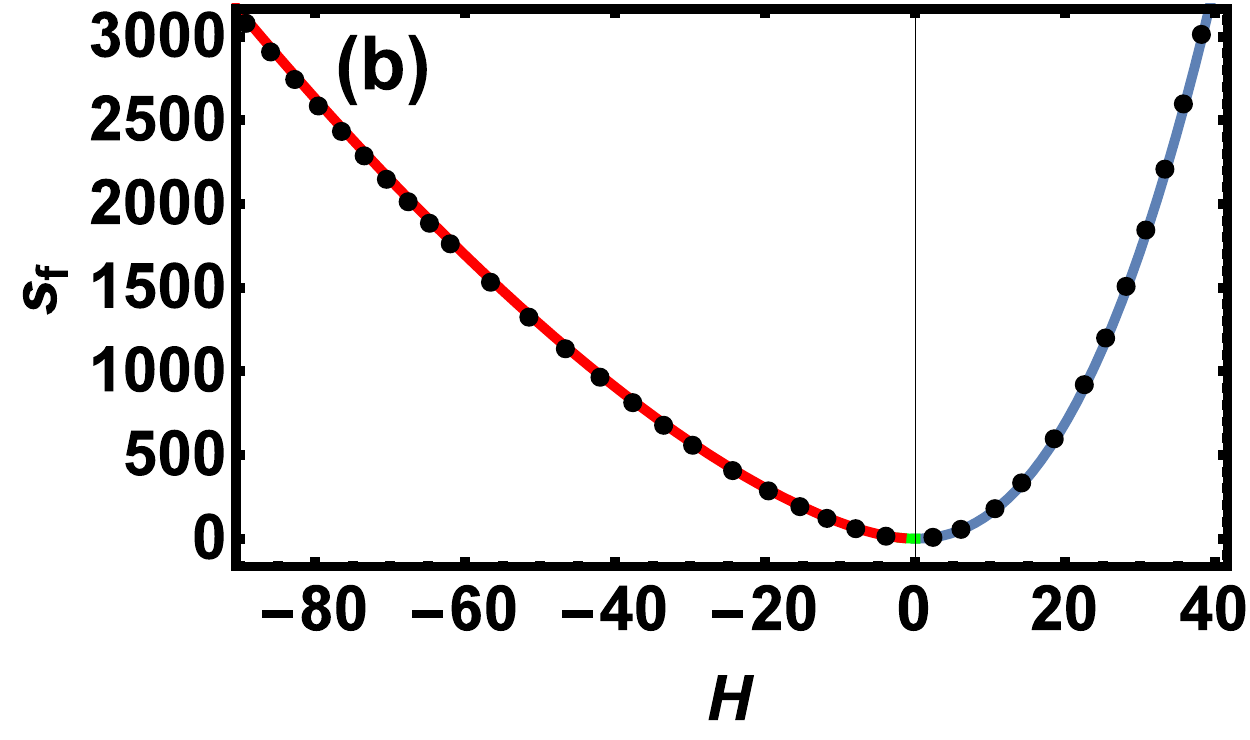}
\caption{Solid line: The rescaled large-deviation function $s_{\text{f}}(H)$, defined by Eqs.~(\ref{actiondgen}) and~(\ref{Svss}), for the flat initial condition. The analytical form of $s_{\text{f}}(H)$ is given by Eqs.~(\ref{eq:Psi_def})-(\ref{eq:sF_exact}). Shown are the regimes of typical fluctuations (a) and large deviations (b). The different intervals $J_{1,2,3}$, see Eq.~(\ref{eq:Jdefs}), are plotted in different colors, but $s_{\text{f}}(H)$ is an analytic function of $H$.  Symbols: numerical results from Ref. \citep{MKV}.}
\label{fig:sF}
\end{figure}

The remainder of this paper is organized as follows. To make this paper self-contained, we present in section \ref{sec:OFM} the OFM formulations for the flat \cite{KK2007,MKV} and stationary \cite{Janas2016} interfaces.
In section \ref{sec:symmetries} we discuss symmetries of the KPZ equation, as manifested in the OFM.
In section \ref{sec:mapping} we exploit a time-reversal symmetry in order to find an exact mapping between the OFM problems for the flat and stationary interfaces, and obtain $S_{\text{f}}(H)$ exactly.
We summarize and discuss our results in Sec. \ref{disc}.
A more detailed discussion of the time-reversal symmetry is relegated to the Appendix.

\section{Optimal fluctuation method}

\label{sec:OFM}

We will use the subscripts  `f' and `st' for the flat and stationary initial conditions, respectively. We will omit the subscript in equations which are valid for both initial conditions. Let us start from the flat interface.

\subsection{Flat interface}

We introduce the time $T$ at which the interface height is measured, so $H=h\left(x=0,t=T\right)$.
After rescaling  $x/\sqrt{\nu T}\to x$, $t/T \to t$, $\left|\lambda\right|h/\nu\to h$, Eq.~(\ref{eq:KPZ_dimensional}) takes the dimensionless form \citep{MKV}
\begin{equation}
\label{eq:KPZ_dimensionless}
\partial_{t}h=\partial_{x}^{2}h-\frac{1}{2}\left(\partial_{x}h\right)^{2}+\sqrt{\epsilon} \, \xi\left(x,t\right),
\end{equation}
where $\epsilon=D\lambda^{2}\sqrt{T}/\nu^{5/2}$ is the dimensionless noise strength, and we assume $\lambda<0$ without loss of generality \citep{signlambda}.
In the weak-noise (or short-time) limit, which formally corresponds to $\epsilon \to 0$, one can use the saddle-point approximation in order to evaluate
the proper path integral of Eq.~(\ref{eq:KPZ_dimensionless}). This leads to
a minimization problem for the action $s_{\text{dyn}}$, where
\begin{equation}
\label{eq:sdyn_def}
s_{\text{dyn}}=\frac{1}{2}\int_{0}^{1}dt\int_{-\infty}^{\infty}dx\left[\partial_{t}h-\partial_{x}^{2}h+\frac{1}{2}\left(\partial_{x}h\right)^{2}\right]^{2}.
\end{equation}
It is convenient to recast the ensuing Euler-Lagrange equation into Hamiltonian equations for the optimal history $h\left(x,t\right)$ of the height profile and its canonically conjugate ``momentum'' $\rho\left(x,t\right)$ which describes the optimal realization of the noise $\xi$ \citep{Fogedby1998, KK2007, MKV}:
\begin{eqnarray}
  \partial_{t} h &=& \delta \mathcal{H}/\delta \rho = \partial_{x}^2 h - \frac{1}{2} \left(\partial_x h\right)^2+\rho ,  \label{eqh}\\
  \partial_{t}\rho &=& - \delta \mathcal{H}/\delta h = - \partial_{x}^2 \rho - \partial_x \left(\rho \partial_x h\right). \label{eqrho}
\end{eqnarray}
Here
\begin{equation*}
\mathcal{H}=\int_{-\infty}^{\infty}\!dx\,\rho\left[\partial_{x}^{2}h- \frac{1}{2}\left(\partial_{x}h\right)^{2}+\rho/2\right]
\end{equation*}
 is the Hamiltonian.

The initial condition for the flat interface is
\begin{equation}
\label{eq:flat_IC}
h_{\text{f}}\left(x,t=0\right)=0.
\end{equation}
The constraint
\begin{equation}
\label{eq:h_1_H}
h(x=0,t=1)=H
\end{equation}
 leads to
\begin{equation}
\label{pT}
\rho\left(x,t=1\right)=\Lambda\,\delta\left(x\right),
\end{equation}
where $\Lambda $ is a Lagrange multiplier whose value is ultimately determined by $H$.

Having solved the OFM problem, one can evaluate the rescaled action $s_{\text{f}}=s_{\text{dyn}}$, which is nothing but the short-time large-deviation function of the height:
\begin{equation}
\label{eq:sdyn_recast}
s_{\text{dyn}}=\frac{1}{2}\int_{0}^{1}dt\int_{-\infty}^{\infty}dx\,\rho^{2}\left(x,t\right).
\end{equation}
This gives $\mathcal{P}$ up to logarithmic accuracy: $-\ln \mathcal{P} \simeq s/\epsilon$, or
\begin{equation}
-\ln\mathcal{P}\left(H,T\right)\simeq\frac{\nu^{5/2}}{D\lambda^{2}\sqrt{T}}\,\,s\left(\frac{\left|\lambda\right|H}{\nu}\right)
 \label{actiondgen}
\end{equation}
in the physical variables.
$s(H)$ is the dimensionless large-deviation function of the short-time height distribution.
Its relation to $S(H)$, which appears in the Abstract and Introduction, is
\begin{equation}
\label{Svss}
S\left(H\right)=\frac{\nu^{5/2}}{D\lambda^{2}}s\left(\frac{\left|\lambda\right|H}{\nu}\right).
\end{equation}

\subsection{Stationary interface}

The effective action for the stationary initial condition is a sum of two terms:
$s_{\text{st}}=s_{\text{dyn}} + s_{\text{in}}$, where $s_{\text{dyn}}$ (\ref{eq:sdyn_def})
is the dynamic contribution, and
\begin{equation}
\label{cost}
s_{\text{in}}=\int_{-\infty}^{\infty}dx\left.\left(\partial_{x}h\right)^{2}\right|_{t=0}
\end{equation}
is the ``cost'' of the (\emph{a priori} unknown) initial height profile.
As a result, the initial condition~(\ref{eq:flat_IC}) gives way to  \citep{Janas2016}
\begin{equation}
\label{eq:OFM_initial_condition_st}
\rho_{\text{st}}\left(x,t=0\right)+2\partial_{x}^{2}h_{\text{st}}\left(x,t=0\right)=\Lambda\delta\left(x\right).
\end{equation}
In addition, we demand $\partial_{x}h_{\text{st}}\left(\left|x\right|\to\infty,0\right)\to0$, guaranteeing the boundedness of $s_{\text{in}}$.
Finally, without losing generality, we can pin the initial Brownian interface at $x = 0$:
\begin{equation}
\label{eq:hB_0_0}
h_{\text{st}}\left(x=0,t=0\right)=0.
\end{equation}

\section{KPZ Symmetries in the OFM formalism}

\label{sec:symmetries}

The KPZ equation respects several symmetries, which play an important role in its analysis \citep{KPZ, Frey1996, Canet2011, Mathey2017, HHZ, Barabasi, Krug, Takeuchi2017}. How do these symmetries manifest themselves in the OFM equations~(\ref{eqh}) and~(\ref{eqrho})? First of all, the KPZ equation is invariant under translations of $x$, $t$ or $h$, leading to an invariance of the OFM equations under the transformation
\begin{eqnarray}
\label{eq:h_translation}
h\left(x-x_{0},t-t_{0}\right)+C &\to& h\left(x,t\right),\\
\rho\left(x-x_{0},t-t_{0}\right) &\to& \rho\left(x,t\right)
\end{eqnarray}
for arbitrary $x_0$, $t_0$ and $C$.
Secondly, there is invariance under spatial inversion
\begin{equation}
\label{eq:spatial_inversion}
h\left(-x,t\right)\to h\left(x,t\right),\quad\rho\left(-x,t\right)\to\rho\left(x,t\right).
\end{equation}
Thirdly, there is invariance under the Galilean transformation
\begin{equation}
xv+h\left(x-vt,t\right)\to h\left(x,t\right),\quad\rho\left(x-vt,t\right)\to\rho\left(x,t\right)
\end{equation}
for arbitrary $v$.
The next two symmetries are nontrivial and involve time reversal.
The OFM equations are invariant under the transformation \citep{Canet2011, footnote:HopfCole}
\begin{eqnarray}
\label{eq:h_time_reversal1}
-h\left(x,-t\right)-2\ln\left|2 \rho\left(x,-t\right)\right|&\to&h\left(x,t\right),\\
\label{eq:rho_time_reversal1}
\rho\left(x,-t\right)&\to&\rho\left(x,t\right).
\end{eqnarray}
The other nontrivial symmetry is the following \citep{Frey1996, Canet2011, Mathey2017}:
\begin{eqnarray}
\label{eq:h_time_reversal2}
-h\left(x,-t\right)&\to&h\left(x,t\right),\\
\label{eq:rho_time_reversal2}
\rho\left(x,-t\right)+2\partial_{x}^{2}h\left(x,-t\right)&\to&\rho\left(x,t\right).
\end{eqnarray}

It is important to note that, while the symmetries~(\ref{eq:h_translation})-(\ref{eq:rho_time_reversal1}) can be extended  to arbitrary spatial dimension, the last symmetry~(\ref{eq:h_time_reversal2}) and~(\ref{eq:rho_time_reversal2}) only holds in 1+1 dimension. It is intimately related to the simple, $\lambda$-independent form of the stationary distribution of interface profiles [the latter determines the ``cost"~(\ref{cost}) of the initial interface in terms of the action].
Technically, the symmetry~(\ref{eq:h_time_reversal2}) and~(\ref{eq:rho_time_reversal2}) is  a consequence of the following exact property of the KPZ action.
For any given profiles $h_0(x)$ and $h_1(x)$ which satisfy $\partial_{x}h_{0}\left(\left|x\right|\to\infty\right)=\partial_{x}h_{1}\left(\left|x\right|\to\infty\right)=0$, and for any (not necessarily optimal) trajectory $h(x,t)$, satisfying $h(x,0)=h_0(x)$ and $h(x,1)=h_1(x)$, one has
\begin{eqnarray}
\label{eq:symmetry_trajectorywise_action}
&&\antiquad \!\! s_{\text{in}}\left[h_{0}\left(x\right)\right]+s_{\text{dyn}}\left[h\left(x,t\right)\right] \nonumber\\
&&\qquad \qquad \; = s_{\text{in}}\left[-h_{1}\left(x\right)\right]+s_{\text{dyn}}\left[-h\left(x,1-t\right)\right].
\end{eqnarray}
Equations~(\ref{eq:h_time_reversal2})-(\ref{eq:symmetry_trajectorywise_action}) are vital ingredients in the mapping, performed in the next section.
In the Appendix we prove Eq.~(\ref{eq:symmetry_trajectorywise_action}) and
show that the symmetry~(\ref{eq:h_time_reversal2}) and~(\ref{eq:rho_time_reversal2}) follows from it.

\section{Mapping between flat and stationary problems}
\label{sec:mapping}

\subsection{Relation between optimal histories}

The symmetries of the previous section are in general violated by the boundary conditions of the OFM problem.
The stationary interface provides a remarkable exception. Here the whole problem is invariant under a combination of the symmetry~(\ref{eq:h_time_reversal2}) and~(\ref{eq:rho_time_reversal2}) and proper translations of $t$ and $h$. Indeed, the boundary conditions~(\ref{eq:h_1_H}),~(\ref{pT}),~(\ref{eq:OFM_initial_condition_st}) and~(\ref{eq:hB_0_0}) are invariant under the transformation
\begin{eqnarray}
\label{eq:h_st_transformation}
&&H-h_{\text{st}}\left(x,1-t\right)\to h_{\text{st}}\left(x,t\right),\\
&&\rho_{\text{st}}\left(x,1-t\right)+2\partial_{x}^{2}h_{\text{st}}\left(x,1-t\right)\to\rho_{\text{st}}\left(x,t\right).
\end{eqnarray}
Less surprisingly, the boundary conditions are also invariant under spatial inversion~(\ref{eq:spatial_inversion}) \citep{Janas2016}.

For the stationary interface, there is a unique solution to the OFM problem at subcritical heights $H > H_c =-3.70632489 \dots$ \citep{Janas2016, LeDoussal2017}, so this solution must respect all of the problem's symmetries. In particular,
$h_{\text{st}}\left(x,t\right)$ must obey the equations
\begin{equation}
\label{eq:symmetric_solutions_ell_0}
h_{\text{st}}\left(x,t\right)=h_{\text{st}}\left(-x,t\right)=H-h_{\text{st}}\left(x,1-t\right).
\end{equation}
At supercritical heights $H < H_c$ there are \emph{three} solutions to the OFM problem. There is a non-optimal (and therefore, non-physical) solution which satisfies the symmetries~(\ref{eq:symmetric_solutions_ell_0}), and whose action coincides with the analytic branch $s_{\text{st}}^{\text{a}}\left(H\right)$ of the large-deviation function $s_{\text{st}}(H)$.
In addition, there are two symmetry-broken, optimal solutions with equal actions, corresponding to the non-analytic branch of $s_{\text{st}}(H)$ \citep{Janas2016, LeDoussal2017}.
These two solutions do not satisfy Eq.~(\ref{eq:symmetric_solutions_ell_0}), but they do satisfy the \emph{combined} symmetry $h_{\text{st}}\left(x,t\right)=H-h_{\text{st}}\left(-x,1-t\right)$ \citep{SKM2018,Janasnum1}.

Henceforth, we will only consider solutions $h_{\text{st}}(x,t)$ which satisfy the symmetries~(\ref{eq:symmetric_solutions_ell_0}), and denote them with the superscript `a' (from the word ``analytic"). These solutions are optimal among all the histories $h(x,t)$ which are conditioned by $h(x=0,t=1)=H$ and respect spatial mirror symmetry. As one can see from Eq.~(\ref{eq:symmetric_solutions_ell_0}), for these solutions
\begin{equation}
\label{eq:h_t_and_H_half}
h_{\text{st}}^{\text{a}}\left(\!x,\frac{1}{2}\right)=\frac{H}{2},
\end{equation}
that is, the interface $h_{\text{st}}^{\text{a}}\left(x,t\right)$ is \emph{flat} at $t=1/2$ \cite{Janasnum2}.
Now,  the OFM problem for the flat interface is known to have a unique solution, which respects spatial mirror symmetry \citep{KK2007, KK2008, KK2009, MKV}. Therefore, it follows from Eq.~(\ref{eq:h_t_and_H_half})
that $h_{\text{st}}^{\text{a}}\left(x,1/2\le t\le1\right)$ is the optimal interface history which leads from a flat interface at height $H/2$ at $t=1/2$ to height $H$ at $x=0$ and $t=1$.
In its turn, this implies a relation between the solutions to the OFM problems for the stationary and flat interfaces:
\begin{equation}
\label{eq:h_S_h_F}
\!\! \frac{H}{2}+h_{\text{f}}\left( \!\! \sqrt{2}\,x,2t-1,\frac{H}{2}\right) \! = h_{\text{st}}^{\text{a}}\left( \! x,\frac{1}{2} \le t \le 1,H  \!  \right).
\end{equation}
Here the dependence of the profiles on $H$ is indicated explicitly, by adding the third argument to $h$.
The factor $\sqrt{2}$ in Eq.~(\ref{eq:h_S_h_F}) comes from the same rescaling of units which leads to Eq.~(\ref{eq:KPZ_dimensionless}).
Using Eqs.~(\ref{eqrho}) and~(\ref{eq:h_S_h_F}), we obtain an additional relation,
\begin{equation}
\label{eq:rhos_connection}
2\rho_{\text{f}}\left(\sqrt{2}\,x,2t-1,\frac{H}{2}\right) = \rho_{\text{st}}^{\text{a}}\left(x,\frac{1}{2}\le t\le1,H\right),
\end{equation}
between the optimal realizations of the noise $\rho_{\text{f}}\left(x,t\right)$ and $\rho_{\text{st}}^{\text{a}}\left(x,t\right)$ which correspond to the interface histories $h_{\text{f}}\left(x,t\right)$ and $h_{\text{st}}^{\text{a}}\left(x,t\right)$, respectively.
If $h_{\text{st}}^{\text{a}}\left(x,t\right)$ is known, $h_{\text{f}}\left(x,t\right)$ can be obtained directly from Eq.~(\ref{eq:h_S_h_F}). Conversely, if $h_{\text{f}}\left(x,t\right)$ is known, $h_{\text{st}}^{\text{a}}\left(x,1/2 \le t\le1\right)$ is found from Eq.~(\ref{eq:h_S_h_F}). Then, using Eq.~(\ref{eq:symmetric_solutions_ell_0}), one can obtain $h_{\text{st}}^{\text{a}}\left(x,0\le t\le1/2\right)$ as well.

\subsection{Relation between height distributions}

Now we can use the relations~(\ref{eq:h_S_h_F}) and~(\ref{eq:rhos_connection}) between the solutions to the OFM problems for the flat and stationary interfaces in order to obtain a relation between their actions.
Plugging $t=0$ into Eq.~(\ref{eq:symmetric_solutions_ell_0}) we find
\begin{equation}
\label{eq:hB_t_inversion}
h_{\text{st}}^{\text{a}}\left(x,t=0\right)=H-h_{\text{st}}^{\text{a}}\left(x,t=1\right).
\end{equation}
By virtue of Eq.~(\ref{eq:h_t_and_H_half})
\begin{equation}
\label{eq:flat_cost_vanishes}
s_{\text{in}}\left[h_{\text{st}}^{\text{a}}\left(x,t=1/2\right)\right] = 0
\end{equation}
(the ``cost'' of the flat interface is zero). Let us now write $s_{\text{st}}^{\text{a}} (H) = s_{\text{st},1}+s_{\text{st},2}$,  where
\begin{eqnarray}
\antiquad s_{\text{st},1}&=&s_{\text{in}}\left[h_{\text{st}}\left(x,0\right)\right]+\frac{1}{2}\int_{0}^{1/2}\!\!\!dt\int_{-\infty}^{\infty}\!\!\!dx\left[\rho_{\text{st}}^{\text{a}}\left(x,t\right)\right]^{2},\\
\label{eq:st2def}
\antiquad s_{\text{st},2}&=&\frac{1}{2}\int_{1/2}^{1}dt\int_{-\infty}^{\infty}dx\,\left[\rho_{\text{st}}^{\text{a}}\left(x,t\right)\right]^{2}.
\end{eqnarray}
It follows from Eqs.~(\ref{eq:symmetry_trajectorywise_action}), ~(\ref{eq:hB_t_inversion}) and ~(\ref{eq:flat_cost_vanishes}) that $s_{\text{st},1} = s_{\text{st},2}$.
On the other hand, by plugging Eq.~(\ref{eq:rhos_connection}) into  Eq.~(\ref{eq:st2def}),
one can show that $s_{\text{st},2} = \sqrt{2}\,s_{\text{f}}\left(H/2\right)$. Altogether this leads to the simple relation
\begin{equation}
\label{eq:s_St_s_F}
s_{\text{f}}\left(H\right)=\frac{s_{\text{st}}^{\text{a}}\left(2H\right)}{2\sqrt{2}}
\end{equation}
between the large-deviation functions of the flat interface and the analytic branch of the  large-deviation function of the stationary interface.
As mentioned above, $s_{\text{st}}^{\text{a}}(H)$ was found exactly in Ref. \citep{LeDoussal2017}. Using their results and  Eq.~(\ref{eq:s_St_s_F}), we will now present $s_{\text{f}}(H)$ \cite{footnote:comparisonLeDoussal}.

\subsection{$s_{\text{f}}(H)$}

Using similar notation to that in Ref. \citep{LeDoussal2017}, we define
\begin{eqnarray}
\label{eq:Psi_def}
\! \Psi_{0}\left(z\right)&=&\frac{1}{\pi}\int_{0}^{\infty} \! dy\left(1+\frac{1}{y}\right)\sqrt{y}\,\ln\left(1+\frac{ze^{-y}}{y}\right),\\
\! \Phi_{0}\left(z\right)&=&\Psi_{0}\left(z\right)-2z\Psi_{0}'\left(z\right),
\end{eqnarray}
and denote the following intervals:
\begin{eqnarray}
&&\antiquad I_{1}=\left[0,+\infty\right],\quad I_{2}=I_{3}=\left[0,e^{-1}\right], \\
\label{eq:Jdefs}
&&\antiquad J_{1}=\left[0,+\infty\right],\quad J_{2}=\left[\frac{H_{c}}{2},0\right],\quad J_{3}=\left[-\infty,\frac{H_{c}}{2}\right] .
\end{eqnarray}
To remind the reader, $H_c<0$ in our units.  The large-deviation function $s_{\text{f}}(H)$ is given in a parametric form by the following
equations:
\begin{eqnarray}
\label{eq:sF_exact0}
&& e^{-H}\!= \! \begin{cases}
\! z\left[\Psi_{0}'\left(z\right)\right]^{2}, & z\in I_{1},\;H\in J_{1},\\
\! z\left[\Psi_{0}'\left(z\right)-\frac{2}{z}\sqrt{-W_{0}\left(-z\right)}\right]^{2} \!\!, & z\in I_{2},\;H\in J_{2},\\
\! z\left[\Psi_{0}'\left(z\right)-\frac{2}{z}\sqrt{-W_{-1}\left(-z\right)}\right]^{2} \!\!, & z\in I_{3},\;H\in J_{3},
\end{cases} \nonumber\\\\
\label{eq:sF_exact}
&&\frac{s_{\text{f}}}{2\sqrt{2}}=\begin{cases}
\Phi_{0}\left(z\right), & z\in I_{1},\\
\Phi_{0}\left(z\right)+\frac{4}{3}\left[-W_{0}\left(-z\right)\right]^{3/2}, & z\in I_{2},\\
\Phi_{0}\left(z\right)+\frac{4}{3}\left[-W_{-1}\left(-z\right)\right]^{3/2}, & z\in I_{3},
\end{cases}
\end{eqnarray}
where $W_{0}(\dots)$ and $W_{-1}(\dots)$ are the first and second real-valued branches of the Lambert function, respectively \citep{Lambert1996, Lambert_wolfram}. $W(z)$ is defined implicitly as the root of the equation $W e^W = z$. The resulting
$s_{\text{f}}(H)$ is plotted in Fig.~\ref{fig:sF}. It is seen to agree perfectly with the numerical results of Ref. \citep{MKV}.

In spite of the presence of three branches in Eqs.~(\ref{eq:sF_exact0}) and (\ref{eq:sF_exact}),
$s_{\text{f}}(H)$ is an analytic function, as is $s_{\text{st}}^{\text{a}}\left(H\right)$ \citep{LeDoussal2017}.
The asymptotic behaviors of $s_{\text{f}}(H)$ are
\begin{equation}
s_{\text{f}} \! \left(H\right)\!=\!\begin{cases}
\frac{8\sqrt{2}}{15\pi}H^{5/2}+\frac{8\sqrt{2}}{3\pi}H^{3/2}\ln H\\
+\frac{4\sqrt{2}}{9\pi}\left[2+3\ln\left(\frac{4}{9\pi^{2}}\right)\right]H^{3/2}+\dots, & H\to+\infty,\\
\sqrt{\!\frac{\pi}{2}}\,H^{2}+\sqrt{\!\frac{\pi}{72}}\,\left(\pi-3\right)H^{3}+\dots, & \left|H\right|\ll1,\\
\frac{8\sqrt{2}}{3}\!\left|H\right|^{3/2}\!\!-\!8\sqrt{2}\,\ln\!\left(2\right)\!\left|H\right|^{1/2}\!+\dots , & H\to-\infty .
\end{cases}
\end{equation}
The leading-order terms of the both tails $|H|\to \infty$ were obtained in Refs.~\cite{KK2007, KK2008, KK2009, MKV}. The subleading terms have been previously unknown.
As has been known for some time \cite{KK2007,MKV}, the tail $H \to -\infty$ agrees with the slower-decaying tail of the GOE Tracy-Widom distribution \citep{TracyWidom1996}.

Using Eqs.~(\ref{actiondgen}), (\ref{eq:sF_exact0}) and (\ref{eq:sF_exact}),
one can evaluate the short-time cumulants
of $\mathcal{P}_{\text{f}}\left(H,t\right)$. There is a shortcut, however, in the form of a simple relation between the
the $q$-th cumulants, $\kappa_{q,\text{f}}$ and $\kappa_{q,\text{st}}$,  for the flat and stationary interfaces, respectively:
\begin{equation}
\label{eq:cumulants_f_and_st}
\kappa_{q,\text{f}}= 2^{\frac{q-3}{2}}\kappa_{q,\text{st}}.
\end{equation}
Plugging the first several nontrivial cumulants $\kappa_{q,\text{st}}$ from the main text and Supplemental Material of Ref.~\citep{LeDoussal2017} into Eq.~(\ref{eq:cumulants_f_and_st}), we find their counterparts for the flat interface:
\begin{eqnarray}
\label{eq:kappa2}
\kappa_{2,\text{f}}&\simeq& D\sqrt{\frac{T}{2\pi\nu}},\\
\kappa_{3,\text{f}}&\simeq&\frac{\left(\pi-3\right)D^{2}\lambda T}{4\pi\nu^{2}}, \\
\label{eq:kappa4}
\kappa_{4,\text{f}}&\simeq&
\frac{\left[5+\left(\sqrt{2}-3\right)\pi\right]\lambda^{2}D^{3}T^{3/2}}{\left(2\pi\right)^{3/2}\nu^{7/2}},\\
\kappa_{5,\text{f}}&\simeq&-\frac{5\left[21+2\left(4\sqrt{2}\,-9\right)\pi\right]\lambda^{3}D^{4}T^{2}}{16\pi^{2}\nu^{5}},\\
\kappa_{6,\text{f}}&\simeq&\! \frac{3\left[252+140 \! \left(\! \sqrt{2}-2\right) \! \pi+\left(15-20\sqrt{2}+8\sqrt{3}\right) \! \pi^{2}\right]}{16\sqrt{2}\,\pi^{5/2}}\nonumber\\&\times&\frac{\lambda^{4}D^{5}T^{5/2}}{\nu^{13/2}} .
\end{eqnarray}
Equation~(\ref{eq:cumulants_f_and_st}) for the particular case $q=2$, and Eq.~(\ref{eq:kappa2}), have been known for a long time \citep{Krug1992}.
Equations~(\ref{eq:kappa2})-(\ref{eq:kappa4}) are in full agreement with Ref. \cite{Gueudre}.
The cumulants $\kappa_{2,\text{f}}$ and $\kappa_{3,\text{f}}$ were also calculated in Ref. \cite{MKV}. The cumulants   $\kappa_{5,\text{f}}$ and $\kappa_{6,\text{f}}$ have not been previously known. Interestingly, the third cumulants of the two distributions coincide.

\section{Summary and Discussion}

\label{disc}

We obtained the exact large-deviation function $s_{\text{f}} (H)$
which describes the  short-time height distribution of the 1d KPZ interface with a flat initial condition, see Eqs.~(\ref{eq:Psi_def})-(\ref{eq:sF_exact}) and Fig.~\ref{fig:sF}.
We achieved this by establishing an exact relation~(\ref{eq:s_St_s_F}) between $s_{\text{f}} (H)$ and the analytic (non-optimal) branch of the short-time large-deviation function $s_{\text{st}} (H)$ of the stationary initial condition. The latter was recently found exactly \citep{LeDoussal2017}.
The relation~(\ref{eq:s_St_s_F}) is a consequence of a time-reversal symmetry of the OFM formulation for the 1d KPZ equation.

It was predicted, for a whole class of initial conditions \citep{MKV,KMSparabola,Janas2016,
MeersonSchmidt2017}, that the OFM results for the $\lambda H\to+\infty$ tail, and sufficiently far into the $\lambda H\to-\infty$ tail, are valid at \emph{all} times. For the droplet initial condition, these predictions are by now firmly established: analytically \cite{LDlongtime,SMP,KLD2}, numerically \cite{SMP,Hartmann}, and rigorously \cite{CorwinGhosal}.

The connection~(\ref{eq:s_St_s_F})  between the large-deviation functions for the stationary and flat interfaces can be extended to finite systems, and in the reciprocal direction. Indeed,
recently the short-time single-point height distribution $\mathcal{P}_{\text{f}}\left(H,L,t\right)$ for an initially flat KPZ interface on a ring of length $2L$ was found in several limits \citep{SMS2018}.
For an initially stationary interface on the same ring the height distribution $\mathcal{P}_{\text{st}}\left(H,L,t\right)$ is presently unknown, but can be found using the results of Ref. \citep{SMS2018}.  Within the regime of parameters where the optimal history for the stationary interface satisfies the symmetries~(\ref{eq:symmetric_solutions_ell_0}) (that is, in the absence of dynamical phase transition), Eq.~(\ref{eq:s_St_s_F}) yields the relation
\begin{equation}
\label{eq:Pstationary_Pflat_ring}
-\ln\mathcal{P}_{\text{st}}\left(H,L,t\right)\simeq-2\ln\mathcal{P}_{\text{f}}\left(\frac{H}{2},L,\frac{t}{2}\right)
\end{equation}
between the two (short-time) distributions.

Finally, the OFM predicts a very simple connection between \emph{any} full-space ($|x| < \infty$) problem with spatial mirror symmetry of the optimal path on the one hand  and the corresponding half-space ($x \ge 0$) problem with the same initial condition and the ``reflecting wall"
boundary condition $\partial_{x}h\left(x=0,t\right)=0$ on the other hand.
Indeed, for the half-space problem, the lower limit of the spatial integration in Eq.~(\ref{eq:sdyn_recast}) is replaced by $0$. Therefore, if $\rho\left(x,t\right)$ in the full-space problem is spatially-symmetric at all $t$,
the large-deviation functions $s\left(H\right)$ and $s^{\text{h.s.}}\left(H\right)$ of the full-space and half-space problems, respectively, are related via $s^{\text{h.s.}}\left(H\right)=\frac{1}{2}s\left(H\right)$. The examples of flat and droplet interfaces (and, more generally, of any deterministic interface which is mirror-symmetric around $x=0$) are almost trivial in this respect.  A less trivial, but still simple, example is the stationary interface. Here the optimal history for the full-space problem is symmetry-broken at supercritical $H$. In the half-space problem, however, the solution must come from the analytic branch $s_{\text{st}}^{\text{a}}\left(H\right)$, which
corresponds to spatially-symmetric interface histories.
Evaluating the integrals~(\ref{eq:sdyn_recast}) and~(\ref{cost}) on these histories over the half space $x \ge 0$ leads to $s_{\text{st}}^{\text{h.s.}}\left(H\right)=\frac{1}{2}s_{\text{st}}^{\text{a}}\left(H\right)$.

\section*{ACKNOWLEDGMENTS}

We thank Tal Agranov for useful discussions and acknowledge financial support from the Israel Science Foundation (grant No. 807/16). N.R.S. was supported by the Clore foundation.

\appendix

\section*{Appendix: Proving Eqs.~(\ref{eq:h_time_reversal2})-(\ref{eq:symmetry_trajectorywise_action})}
\renewcommand{\theequation}{A\arabic{equation}}
\setcounter{equation}{0}

Here we discuss the time-reversal symmetry of the KPZ equation in $1+1$ dimension in some detail.
We prove Eq.~(\ref{eq:symmetry_trajectorywise_action}), and then use it in order to prove the symmetry~(\ref{eq:h_time_reversal2}) and~(\ref{eq:rho_time_reversal2}). Let us denote by $s_{\text{LHS}}$ and $s_{\text{RHS}}$ the left and right hand sides of Eq.~(\ref{eq:symmetry_trajectorywise_action}), respectively. Using Eqs.~(\ref{eq:sdyn_def}) and (\ref{cost}), we obtain
\begin{eqnarray}
\label{1}
\antiquad s_{\text{LHS}}&=&\int_{-\infty}^{\infty} \! \left(\partial_{x}h_{0}\right)^{2}dx\nonumber\\
\antiquad &+&\frac{1}{2} \int_{-\infty}^{\infty} \! dx\int_{0}^{1}dt\,\left[\partial_{t}h-\partial_{x}^{2}h+\frac{1}{2}\left(\partial_{x}h\right)^{2}\right]^{2}
\end{eqnarray}
and
\begin{eqnarray}
\label{2}
\antiquad s_{\text{RHS}}&=&\int_{-\infty}^{\infty} \! \left(\partial_{x}h_{1}\right)^{2}dx\nonumber\\
\antiquad &+& \frac{1}{2} \int_{-\infty}^{\infty} \!\!\! dx\int_{0}^{1}dt\,\left[\partial_{t}h+\partial_{x}^{2}h+\frac{1}{2}\left(\partial_{x}h\right)^{2}\right]^{2} .
\end{eqnarray}
Subtracting Eq.~(\ref{1}) from Eq.~(\ref{2}), we obtain
\begin{eqnarray}
\label{eq:app1}
&& \antiquad \!\! s_{\text{RHS}}-s_{\text{LHS}}=\int_{-\infty}^{\infty} \! dx\left[\left(\partial_{x}h_{1}\right)^{2}-\left(\partial_{x}h_{0}\right)^{2}\right]\nonumber\\
&&\qquad +\int_{-\infty}^{\infty} \! dx\int_{0}^{1}dt\,\left[2 \partial_{t}h\partial_{x}^{2}h+\partial_{x}^{2}h\left(\partial_{x}h\right)^{2}\right] .
\end{eqnarray}
Now we use the identity
\begin{equation}
2\partial_{t}h\partial_{x}^{2}h=2\partial_{x}\left(\partial_{t}h\partial_{x}h\right)-\partial_{t}\left[\left(\partial_{x}h\right)^{2}\right]
\end{equation}
in~Eq.~(\ref{eq:app1}) to obtain
\begin{equation}
s_{\text{RHS}}-s_{\text{LHS}}= \!\! \int_{-\infty}^{\infty} \!\!\! dx \! \int_{0}^{1} \! dt\,\partial_{x}\! \left[2\partial_{x}h\partial_{t}h+ \! \frac{1}{3}  \left(\partial_{x}h\right)^{3}\right] \! =0,
\end{equation}
completing the proof of Eq.~(\ref{eq:symmetry_trajectorywise_action}).

The symmetry~(\ref{eq:h_time_reversal2}) and~(\ref{eq:rho_time_reversal2}) follows from Eq.~(\ref{eq:symmetry_trajectorywise_action}).
Indeed, for given profiles $h_0(x)$ and $h_1(x)$, let $h_{*}(x,t)$ be the history which minimizes $s_{\text{dyn}}\left[h\left(x,t\right)\right]$ under the constraints
\begin{equation}
\label{eq:h0_h1}
h\left(x,0\right)=h_{0}\left(x\right),\quad  h\left(x,1\right)=h_{1}\left(x\right).
\end{equation}
Then it follows from Eq.~(\ref{eq:symmetry_trajectorywise_action}) that
\begin{equation}
\label{eq:hd}
\tilde{h}_{*}\left(x,t\right)\equiv-h_{*}\left(x,1-t\right)
\end{equation}
minimizes $s_{\text{dyn}}\left[\tilde{h}\left(x,t\right)\right]$
under the constraints
\begin{equation}
\label{eq:h0_h1_rev}
\tilde{h}\left(x,0\right)=-h_{1}\left(x\right) ,\quad  \tilde{h}\left(x,1\right)=-h_{0}\left(x\right).
\end{equation}

As is seen from Eqs.~(\ref{eq:KPZ_dimensionless}) and~(\ref{eq:hd}), the optimal realizations $\xi_{*}^{\text{opt}}\left(x,t\right)$ and $\tilde{\xi}_{*}^{\text{opt}}\left(x,t\right)$ of the noise, which correspond to the optimal histories $h_{*}\left(x,t\right)$ and $\tilde{h}_{*}\left(x,t\right)$, respectively, are related via
\begin{equation}
\label{eq:xistar}
\tilde{\xi}_{*}^{\text{opt}}\left(x,t\right)=\xi_{*}^{\text{opt}}\left(x,1-t\right)+\frac{2}{\sqrt{\epsilon}}\partial_{x}^{2}h_{*}\left(x,1-t\right).
\end{equation}
Since $h_{*}\left(x,t\right)$ and $\tilde{h}_{*}\left(x,t\right)$ are optimal histories, and in view of the relation $\rho\left(x,t\right)=\sqrt{\epsilon}\,\xi^{\text{opt}}\left(x,t\right)$, which follows from Eqs.~(\ref{eq:KPZ_dimensionless}) and~(\ref{eqh}), we find that Eqs.~(\ref{eq:hd}) and~(\ref{eq:xistar}) lead to the invariance of the OFM equations~(\ref{eqh}) and~(\ref{eqrho}) under the transformation~(\ref{eq:h_time_reversal2}) and~(\ref{eq:rho_time_reversal2}). Of course, this invariance can be also verified directly.


\begin{thebibliography}{99}
\bibitem{signlambda}
Changing the sign of $\lambda$ is  equivalent to changing the sign of $h$.
\end{thebibliography}
\end{document}